\documentclass[11pt,oneside,a4paper]{article}

\usepackage{fancyhdr}
\title{\bf LA-UR-02-3761 \\ \rm The Software Anatomy of a Flexible Accelerator Simulation Engine}
\author{Nicholas D. Pattengale \ \ \ \ \ \ \ \ \ Christopher K. Allen\\ 
\texttt{nickp,ckallen@lanl.gov}
}
\date{June 10, 2002}
\begin{document}
\maketitle

\begin{abstract}
A modular, maintainable and extensible particle beam simulation architecture is presented.  
Design considerations for single particle, multi particle, and rms envelope simulations (in two and three dimensions) are outlined.  
Envelope simulation results have been validated against Trace3D.  
Hybridization with a \it physics-centric \rm contol-system abstraction provides a convenient environment for rapid deployment of applications employing model-reference control strategies.
\end{abstract}

\section{Background}

\subsection{Discovering a Simulation Architecture}

Our group has designed and implemented a unified accelerator Application Programming Interface (API) called XAL\cite{XAL}.
XAL is designed to aid in the development of science control applications for beam physics.
Accordingly, the XAL API is a  \it  physics-centric  \rm  software programming interface.
The physics applications interact with a model of an accelerator that resides in computer memory.  
XAL also contains the software infrastructure that creates the accelerator model.
XAL loads a text-based (XML) description of an accelerator and assembles software objects such that an accurate model of the accelerator exists in computer memory.  XAL is based on UAL \cite{UAL}, the Unified Accelerator Library.

The original motivation for XAL was to provide an accelerator independent interface for applications to interact with I/O from a live accelerator.
This allows physicists to write beam physics control applications (Orbit Correctors, Beam Profile Monitors, RF Tuners, etc.) to the XAL API so that they can run on any accelerator.
Some pseudo-code illustrating the principles of an XAL-based Orbit Correction Application may illustrate the essence of the concept.

\begin{verbatim}
Accelerator  theAccel = XALFactory.newAccel(``sns.xml'')
BPM[]        theBPMs       = theAccel.getNodesOfType(BPM)
HorzDipole[] theCorrectors = theAccel.getNodesOfType(DCH)
for each BPM in theBPMs
      read BPM.avgPos() and set a corrector magnet accordingly
\end{verbatim}

To aid in writing applications that take into account design values, the accelerator description file contains all design information for the accelerator.
This condition allows, for example, a physics application to compare the design field of a quadrupole with its read-back (runtime) field.

With all design information incorporated into a software model of an accelerator, we have discovered an excellent simulation engine.  
As long as the software accelerator has a convenient means for traversing beam-line devices in a spatially sequential manner, we can use design values along the way to simulate beam-dynamics.

This scenario allows for a drastic departure from traditional accelerator simulation codes.  Traditionally simulators have been isolated software products.  They load some type of lattice description of an accelerator and apply predefined beam-dynamics to an initial beam.  Ultimately this design has led to huge codes (to account for various beam-line element types).  Further, these codes typically operate with only one type of simulation (multi-particle or rms envelope, but not both).

The architecture presented here contains a novel approach to the simulation domain.  It is our conjecture that the method presented here better captures reality in that there is some sort of \it software beam \rm actually traversing a software model of a real accelerator.

\subsection{The Architecture}

Our approach is based upon the Element-Algorithm-Probe Design Pattern \cite{EAP}.  
The core concept of this design pattern is the separation of beam-dynamics code from the actual beam-line elements.  
It is desirable to keep the code that corresponds to beam-line elements as simple as possible so that the application writer has a clean interface to a beam-line element.  
The Element-Algorithm-Probe pattern enforces this concept by requiring beam-dynamics code to exist in a separate entity, called an  \tt  IAlgorithm  \rm.

Deferred until runtime is the binding of beam-dynamics to actual beam-line elements.  
This deployment strategy allows for conceptually correct simulations.  
First it is truly modular.  
The three concepts, beam-line elements, beam-dynamics, and the beam are compartmentalized into separate code.  
Second it is truly maintainable.  
To support a new beam type or new beam-line element type does not cause code bloat.  
Finally it is truly extensible.  
Via the mechanism of a Java interface, various beam-dynamics algorithms can be written for the same type of beam-line element and switched at will at runtime.  
Modularity, maintainability, and extensibility provide true power and flexibility to our architecture.

\section{\tt IProbe, IAlgorithm, and IElement\rm}

\subsection{Technology Introduction}

It may help to understand the facets of Java that we exploit in order to implement the Element-Algorithm-Probe Design Pattern.

At the center of the Element-Algorithm-Probe pattern is the concept of a Java  \it  interface.
\rm Essentially, an interface is a contract between a user and an implementor.
The contract says that the implementor of an interface is required to provide an implementation of the methods defined in the interface.

For example, consider the interface

\begin{verbatim}
public interface Thermometer {   
   public double getTemperature();
}
\end{verbatim}

Using this interface, a programmer can assume being able to perform operations on a thermometer no matter how the thermometer actually obtains the temperature.  
This is desirable because a thermometer implementor can change how the temperature is actually obtained (if, say, a new sensor system was installed) without requiring all thermometer users to recompile their code.  

We use the same idea with beam-dynamics code.  
Beam-dynamics reside in files that  \it  implement  \rm  (the computer science term for acknowledging involvement in the contract from the implementors point of view) the  \tt  IAlgorithm  \rm  interface.  
Since the simulation engine knows how to do beam-dynamics calculations solely by interacting with  \tt  IAlgorithm\rm s, it is trivial to swap beam-dynamics algorithms at will.  

The  \tt  IAlgorithm  \rm  interface looks like this.

\begin{verbatim}
public interface IAlgorithm {
   public void propagate(IElement, IProbe);
   public Class legalElementType();
   public Class legalProbeType();
} 
\end{verbatim}
Conceptually, an  \tt  IAlgorithm  \rm  implementor is required provide an implementation of the method  \tt  propagate()  \rm  to modify the the beam (\tt IProbe\rm ) according to the beam dynamics of the the beam-line element (\tt IElement\rm ).

In essence, all that the simulation engine knows about are the three data types (all defined in interfaces)  \tt  IAlgorithm, IProbe,  \rm  and  \tt  IElement.\rm  
The beauty of this design is that there are separate code locations for beam-line elements, beam-dynamics, and the beam itself.

\subsection{Probes}

The \tt IProbe \rm interface should ideally contain the bare minimum information to fully represent a beam.  
Such beam information consists of beam current, beam charge, particle charge, particle rest energy, particle kinetic energy, etc.
Further, since a probe represents the state of the beam at a position in the beam-line, a probe also contains a beam-line-position attribute.
The current \tt IProbe \rm specification serves the purpose of representing a beam for a single particle , particle ensemble, and envelope simulations (in both two and three dimensions).
Figure 1 represents a suitable inheritance hierarchy of probe types to handle the aforementioned simulation types.

\input{epsf}

\begin{figure}
	\centerline{\epsffile{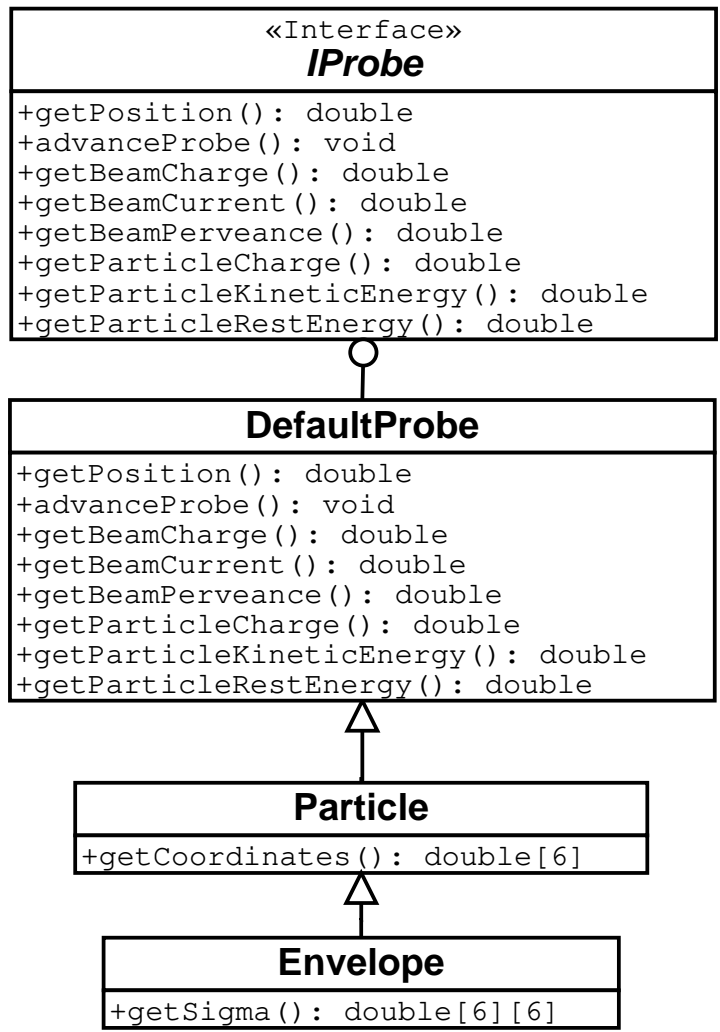}}
	\caption{UML Diagram of Probe Type Hierarchy}
	\label{Figure 1}	
\end{figure}

\subsection{Propagation}

It is important to note that there are various approaches toward simulating beam-dynamics.  
For example, accurate approaches may involve slicing nodes up into small pieces.  
An aggregate of approximations done on sufficiently small elements is typically more accurate than one overall approximation.  
However, normally this is only practical in elements that have special behavior.  
So the question arises:  \it  How are probes propagated through elements?  \rm

It is the responsibility of the  \tt  IAlgorithm  \rm  implementor to handle all beam-dynamics, including the propagation mechanism.  
Sample propagation mechanisms will be presented later in this paper.  
However, keep in mind the most appropriate propagation mechanism when implementing algorithms for the particular problem at hand.

\subsection{Algorithms}

We have already introduced the concept of the  \tt  IAlgorithm  \rm interface.  Now let us pursue a few more details regarding its implementation.

The \tt  IAlgorithm  \rm  interface provides a generic way of assembling algorithms in a simulation engine.  
In practice any particular  \tt  IAlgorithm  \rm implementation only makes sense in the context of a particular beam-line element type and probe type.  
For example, a hypothetical  \tt  IAlgorithm \rm called  \tt  QuadParticleMapper  \rm  would expect a Quadrupole as its  \tt  IElement  \rm  and a Particle as its  \tt  IProbe  \rm.
Providing such specificity is the job of the  \tt  legalElementType()  \rm  and  \tt  legalProbeType()  \rm methods.  
An implementation of the  \tt  QuadParticleMapper  \rm  could look like this.

\begin{verbatim}


public class QuadParticleMapper(IElement p_elem, 
                                IProbe p_probe){
  public Class legalElementType(){return Quadrupole.class;}   
  public Class legalProbeType(){return Particle.class;}
  public void propagate(){Quadrupole/Particle beam dynamics}
}
\end{verbatim}
By providing these methods, the simulation engine can do type checking upon algorithm binding.  
It would not make sense to bind this algorithm to a WireScanner.
Providing these methods helps to avoid that condition.

\section{Design of a Single Particle Simulation}

Designing an actual simulation merely involves putting together Elements, Algorithms, and Probes in a semantically meaningful way.
It turns out that, to the first order, the beam dynamics through a particular node type can be captured by a transfer matrix.
This property allows for a straight-forward means of simulating a particle traveling down a beam-line.
An object-oriented approach would be to create a \tt ParticleMapper \rm class that transforms the \tt Particle \rm probe by the simple vector-matrix multiplication 
$$\vec{\bf{z}}_{n + 1} = \Phi_n \cdot \vec{\bf{z}}_n$$
where $\vec{\bf{z}}_n$ is the coordinate vector of the particle $\left(\begin{array}{cccccc} x & x' & y & y' & z & z' \end{array}\right)^T$ at the start of the node, and $\Phi_n$ is the transfer matrix of the node.
Further, $\Phi_n$ can be obtained by the \tt ParticleMapper \rm via the use of an abstract method that is implemented by beam-dynamics algorithms for individual nodes (QuadrupoleParticleMapper, RFCavityParticleMapper, etc.).
A suitable class design can be seen in Figure 2.

\begin{figure}
	\centerline{\epsffile{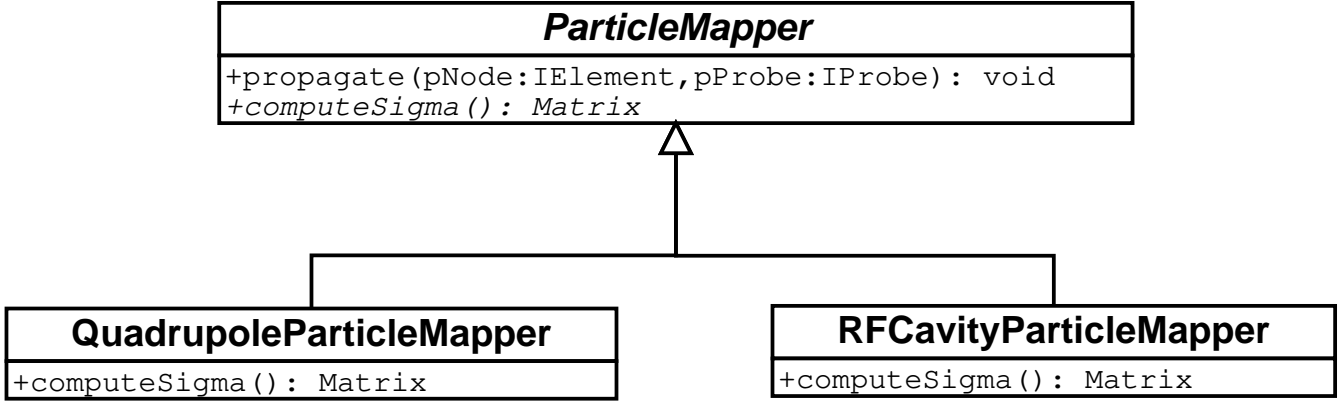}}
	\caption{UML Diagram of ParticleMapper Type Hierarchy}
	\label{Figure 2}	
\end{figure}

To further illustrate some of these concepts, the basic layout of the \tt ParticleMapper \rm class looks like this.

\begin{verbatim}
abstract public Matrix computeTransferMatrix();

public void propagate(IElement pElem, IProbe pProbe)
{
        //type-cast the probe and element to what we expect
        Particle theProbe = ((Particle)pProbe); 
        AcceleratorNode theNode = ((AcceleratorNode)pElem);	

        //do the vector-matrix multiplication
        theProbe.setCoords(computeTransferMatrix
                             .times(theProbe.getCoords()));

        //advance the probe the length of the node
        theProbe.advancePosition(theNode.getLength());
        return;
}
\end{verbatim}

Once the \tt computeTransferMatrix() \rm operations are implemented for the node-specific dynamics, all that remains is writing a driver program.
A driver program binds algorithms to nodes and injects the probe.
Here is a pseudo-code driver\footnote[1]{It is anticipated as a logical extension to these ideas that an \it AlgorithmManager \rm GUI will be written so that simulations can be bound at run-time to the XAL accelerator model and easily hooked into analysis software.}.

\begin{verbatim}

//instantiate the XAL accelerator model
Accelerator  theAccel = XALFactory.newAccel(``sns.xml'')

//bind the algorithms
Quadrupole[] theQuads      = theAccel.getNodesOfType(QUAD)
RFCavity[]   theCavities   = theAccel.getNodesOfType(RFC)
for each QUAD in theQuads
      bind a QuadParticleMapper instance to QUAD
for each RFCav in theCavities
      bind a RFCavityParticleMapper instance to RFCav

//instantiate a probe
Particle p1 = new Particle(initial conditions...)


//run the probe down the beam-line
AcceleratorNode[] theNodes = theAccel.getAllNodes();
for each NODE in theNodes
      NODE.propagate(p1)
\end{verbatim}

And that is it!\footnote[1]{Notice that beam-dynamics between nodes are left out of the demonstration case.  In the actual simulation engine, the \tt propagate() \rm method accounts for space between the position of the probe and the start of the node by calculating the beam-dynamics through a drift space.  However future implementations of XAL may include drift-spaces as an actual node type which would warrant writing a specific \tt DriftSpaceMapper \rm class.}

The particle probe will be transformed by each beam-line element according the the bound algorithm.
Note that the pseudo-code is a basic proof of concept and does not contain the code necessary to broadcast probe increment intermediate data to produce, for example, a plot.

The single particle simulation can be applied to a two-dimensional case by only considering the first four elements of $\vec{\bf{z}}$. 
Further, the single particle simulation can be extended to a multi-particle simulation (in two and three dimensions) by constructing a container of particle probes and writing beam-dynamics algorithms that properly transform the collection.
The only matter that complicates (and complicate it does!) a multi-particle simulation is the concept of space-charge.
Before biting off this task, however, a presentation of another type of simulation that accounts for space-charge is warranted.

\section{Design of an RMS Envelope Simulation}

\subsection{The Concept}

The RMS qualities of a beam can be represented by the 6x6 symmetric matrix $\sigma$  that statistically expresses the boundaries of a beam in transverse, longitudinal, and phase space by using moments of the beam distribution.
RMS Envelopes are convenient because applying beam-dynamics involves a simple matrix operation.  
Namely, the same transfer matrix $\Phi$ used in single particle simulations can propagate rms envelopes according to the conjugation

$$\sigma_{n+1} = \Phi \cdot \sigma_n \cdot \Phi^T$$

The other important concept in this simulation is space charge.  
A $\sigma$ matrix is a statistical representation of a beam, which is a multi-particle entity.  
Therefore, each particle in the beam is aware (electromagnetically) of all other particles in the beam.
It turns out that to the first order the effects of space charge can be captured in a $\Phi$ matrix.
While it may not be mathematically trivial to calculate the matrix, having the calculation in such a form makes the integration into our simulation engine simple.
However it should not be overlooked that this quantity is very important to the correctness of simulation.

\subsection{Propagation}

The envelope simulation is more complex than a single-particle simulation in that we will propagate envelopes through elements using more than one propagation mechanism.  
Specifically, we may be able to compute a better approximation of behavior through quadrupoles than RF Cavities.

This condition is due to the fact that the $\Phi$ matrix for a quadrupole adheres to the semi-group property.

$$\Phi(\Delta s_1 + \Delta s_2) = \Phi(\Delta s_1) \cdot \Phi(\Delta s_2)$$
\centerline{or}
$$\Phi(n \cdot \Delta s) = \Phi^n(\Delta s)$$
where $\Delta s$ is the length of the quadrupole being considered.

To more accurately consider space charge, we take advantage of the semi-group property of the transfer matrices.
In the \tt propagate() \rm method of the \tt SemiGroupEnvelopeMapper \rm  (see Figure 3) we subsection the node (e.g., a quadrupole) into $n$ slices of length $\Delta s = {l \over n}$ where $l$ is the length of the quadrupole.
Then we run the probe through these $n$ slices, applying space charge kicks after every subsection (See Figure 4).

\begin{figure}
	\centerline{\epsffile{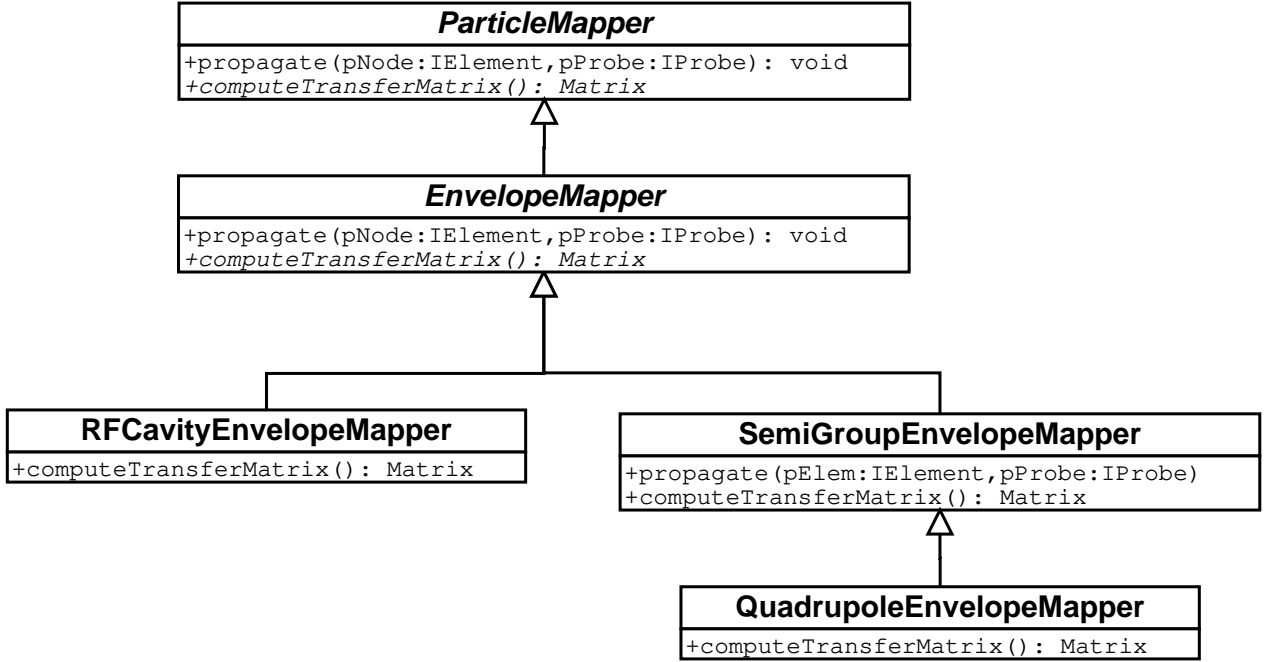}}
	\caption{UML Diagram of EnvelopeMapper Type Hierarchy}
	\label{Figure 3}	
\end{figure}

\begin{figure}
	\centerline{\epsffile{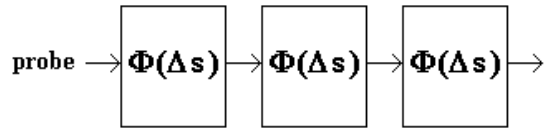}}
	\caption{Graphical Representation of Transformation of RMS Envelope through node with Semi-Group property}
	\label{Figure 4}	
\end{figure}

Since RF Cavity transfer ($\Phi$) matrices do not in general adhere to a semi-group property, we are forced to take a more simplistic approach toward transforming the envelope.  
We will slice the node in two, treating each half as a drift-space (to account for space charge) and hit the envelope in the middle of the node with the numerically approximated $\Phi$ matrix (see Figure 5).

\begin{figure}
	\centerline{\epsffile{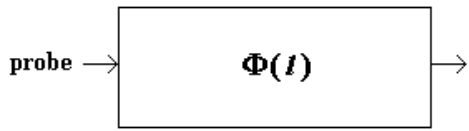}}
	\caption{Graphical Representation of Generic Transformation of RMS Envelope}
	\label{Figure 5}	
\end{figure}

\section{Design of a Particle Ensemble Simulation}

As a final exercise it will be useful to consider the design of a multi-particle simulation.
The true complication of designing a multiple-particle (ensemble) simulation is
 the computation of space-charge effects.
Unfortunately, to model multiple particles, space-charge effects cannot be accurately captured by a transfer matrix.
On the other hand, the architecture outlined in this paper keeps the details of the space-charge calculations from interfering with code cleanliness.

\begin{verbatim}




\end{verbatim}

The two core concepts of a multi-particle (ensemble) simulation are
\begin{itemize}
\item Calculation of the electric self-fields of the ensemble
\item Using the calculated fields to update the particle coordinates.
\end{itemize}
There are various approaches that can be taken for both tasks.
All that we attempt to show here is that by correctly isolating these concepts, a clean software architecture can be maintained.

Namely, an \tt Ensemble \rm probe should encapsulate the logic necessary to obtain the electric self fields of the ensemble.
That being the case, various \tt Ensemble \rm probe implementations could be swapped at will to employ different field calculation techniques.
For example, many electric field calculation techniques involve solving Poisson's equation to obtain the electric potential of the ensemble.
By hiding this code from the simulation engine (it is contained within the \tt Ensemble \rm probe implementation), the implementor could exploit parallel processing facilities (See Figure 6).
\begin{figure}
	\centerline{\epsffile{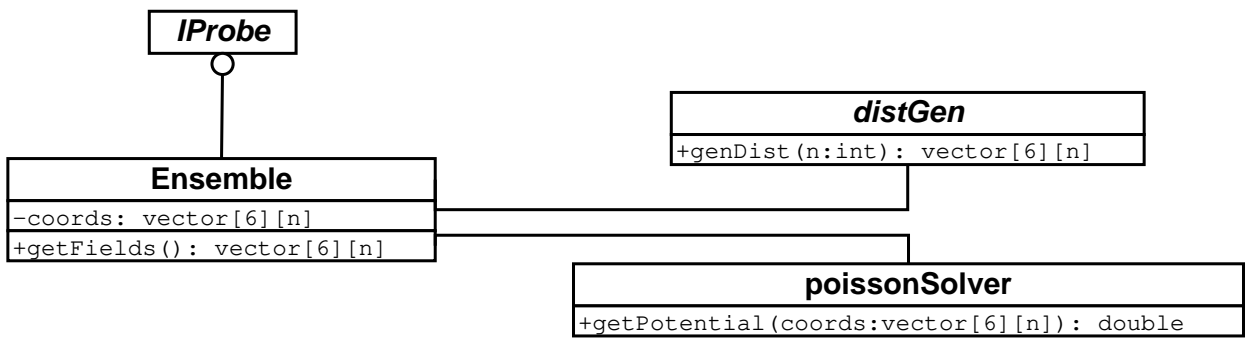}}
	\caption{An \tt Ensemble \rm  probe could vary its implementation details by hiding helper classes.}
	\label{Figure 6}
\end{figure}

By moving the calculation of electric fields out of the beam-dynamics code, the beam-dynamics algorithm developer is free to choose space-charge consideration techniques with minimal impact to code clarity.
One may decide to take the ``thin lens kick'' approach that has been used previously in this paper.
One may alternatively decide to apply a ``trajectory integration'' based approach.
The key point here is that by separating codes into their logical components allows for a high degree of flexibility in simulation technique.

\section{Conclusions and Future Directions}

We are enthusiastic to report that the results obtained in the RMS Envelope Simulation have been validated against Trace3D \cite{Trace3D} Figure 7\footnote[1]{The slight inconsistencies arise from the fact that Trace3D only outputs at the end of nodes whereas our simulation engine produced a higher resolution of intermediate data points.} shows agreement between simulation results of the SNS Medium Energy Beam Transport (MEBT) using both Trace3D and the XAL simulation engine.
It is encouraging that a problem domain with so many interdependencies(particle physics) can be simulated with a clean architecture.

\begin{figure}
	\centerline{\epsffile{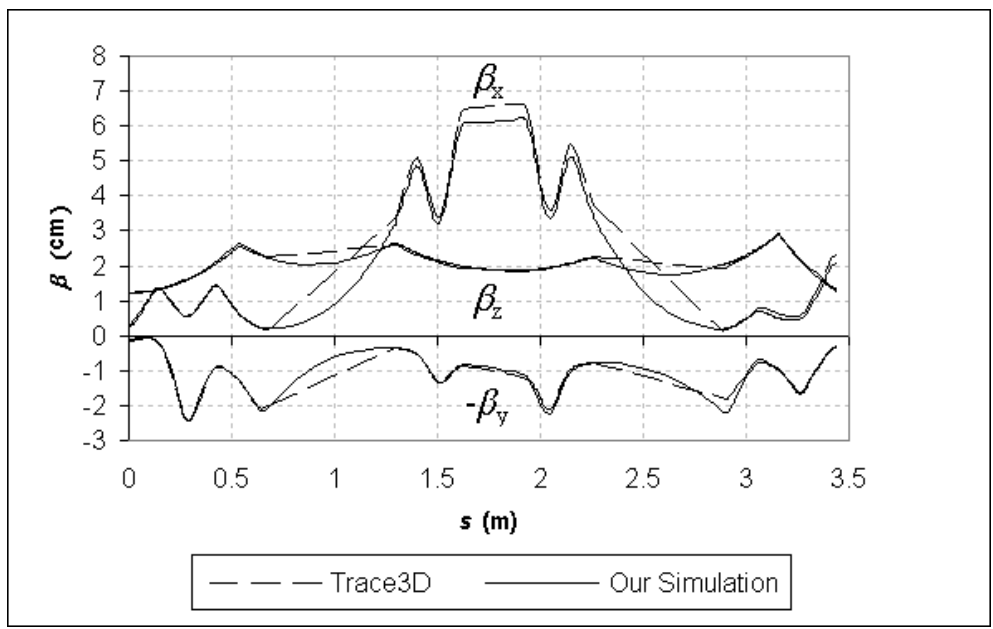}}
	\caption{Plot of Twiss parameter $\beta$ in the SNS MEBT produced by Element-Algorithm-Probe Architecture (solid lines) vs. Trace3D (dashed lines)}
	\label{Figure 7}
\end{figure}

As we move toward the future, we are anticipating the ability to implement model reference control techniques.
That is, within the XAL model there is access to a live accelerator and a simulated accelerator.
Having both at hand allows the comparison of live behavior with simulated behavior to develop control strategies.

The key to effectively implementing an environment conducive to model reference control is architectural discipline when designing both the I/O and simulation aspects of XAL.
As long as the interface to the two are respectively clean, hybridization of the two will be a straightforward extension.

\end{document}